\documentclass[preprint,12pt]{elsarticle}

\usepackage{amssymb}
\usepackage{color}
\usepackage{amsmath}
\usepackage{bm}
\usepackage{subfigure}
\usepackage{float}
\usepackage{hyperref}
\usepackage[capitalize]{cleveref}
\usepackage{booktabs}
\usepackage{mathtools}
\makeatletter
\def\ps@pprintTitle{%
  \let\@oddhead\@empty
  \let\@evenhead\@empty
  \let\@oddfoot\@empty
  \let\@evenfoot\@empty
}
\makeatother

\begin{document}

\begin{frontmatter}

\title{A quantum algorithm for meshless lattice Boltzmann method with unitary interpolation}

\author[inst1]{Chi Zhang}
\author[inst2]{Hefeng Wang}
\author[inst1]{Hua Xiang\corref{cor1}}

\affiliation[inst1]{
  organization={School of Mathematics and Statistics, Wuhan University},
  city={Wuhan},
  country={China}
}

\affiliation[inst2]{
  organization={School of Physics, Xi'an Jiaotong University},
  city={Xi'an},
  country={China}
}

\cortext[cor1]{Corresponding author. Email: hxiang@whu.edu.cn}

\begin{abstract}
  We present a mass-weighted meshless formulation for the shifted incompressible lattice Boltzmann method (LBM) whose streaming step is exactly unitary, combining with the Carleman embedding.
  In traditional meshless LBM, on a scattered node set the semi-Lagrangian streaming step is an interpolation, which is not unitary; a generic block encoding of this matrix therefore requires a normalization proportional to its sparsity and logarithmically many ancillas.
  To overcome the non-unitarity, we instead express the interpolation as the exponential \(\exp(aD_x+bD_y)\) of mimetic derivative operators. Endowing each node with its Voronoi mass and applying a mass-weighted similarity transform renders these derivative operators skew-Hermitian by enforcing moment constraints together with low-wavenumber trigonometric differentiation conditions, so that the interpolation becomes exactly unitary. 
  Such unitarization of interpolation leave the local collision step unchanged up to a rescaling of its quadratic term, where the same weights define the amplitude encoding of the distribution functions.
  For quasi-uniform node sets, the mass-weighting diagonal remains within a unit-order normalization that degrades only mildly as the Carleman truncation order grows, thereby mitigating the cost of block encoding.
  Numerical experiments on the Taylor-Green vortex confirm that the method reproduces the full meshless LBM, with the velocity error decreasing as the Carleman order increases.
\end{abstract}


\end{frontmatter}

\section{Introduction}\label{section 1}
  Computational fluid dynamics~(CFD) remains one of the most demanding areas of scientific computing, and quantum computers have been proposed as a route to accelerating the solution of the partial differential equations underlying incompressible flow. The lattice Boltzmann method~(LBM) is particularly attractive in this direction: it replaces the nonlinear Navier-Stokes system by a kinetic description in which a small set of distribution functions obey a linear streaming step and a local, algebraic collision step \cite{heluo1997}. The locality of the collision operator makes the nonlinearity tractable, while the streaming step is a pure shift on the lattice.

  For a quantum setting, the collision nonlinearity can be addressed by Carleman linearization \cite{carleman1932,kowalski1991}, which embeds the finite-dimensional nonlinear dynamics into an infinite-dimensional linear system on the tensor powers of the state. Truncating this hierarchy at order \(N_C\) yields a finite but high dimensional linear evolution \cite{forets2017,liu2021,krovi2023}. Applying the Carleman linearization to the lattice Boltzmann equation, the entire time step becomes a single linear map on the Carleman vector \(\bm{y}=(\bm{g},\,\bm{g}^{\otimes2},\,\dots,\,\bm{g}^{\otimes N_C})\), so that a collision-and-streaming update reduces to repeated application of a high-dimensional matrix. Several quantum lattice Boltzmann algorithms based on this construction have been  proposed \cite{yepez2001,budinski2022,sanavio2025}, and recent work has analyzed the Carleman truncation error and the block-encoding cost of the full incompressible flow solver, including nontrivial boundary conditions \cite{jennings2025a,jennings2025b}. A central conclusion of these analyses is that the quantum resources are governed by the subnormalization factors of the block-encoded operators, which must be controlled for the algorithm to retain a useful complexity of the quantum linear system solver.

  Classically, the restriction of LBM to uniform Cartesian lattices is a serious practical limitation. To address this limitation, many methods have been developed to lift it by allowing the nodes to move. In the interpolation-supplemented LBM of He, Luo and Dembo the streaming step is performed on a non-uniform mesh by interpolating the populations back onto the reference nodes \cite{heluodembo1996}; the least-squares (meshless) LBM of Shu, Niu and Chew instead constructs  Taylor-expansion based local interpolation and derivative operators on scattered nodes \cite{shu2002}, and radial basis function (RBF) variants have since been developed \cite{strzelczyk2023,fornberg2015}. In all of these the streaming step is written as
  \[
  f_m(X,t+\Delta t)
  =f_m^C(X-\bm e_m\Delta t,t)
  \approx P_m\,\bm f_m^C(X,t),
  \]
  where \(X=\{\bm x_i\}_{i=1}^{N}\) is the scattered node set and \(P_m\) is a translation interpolation matrix that transfers the \(m\)-th population from \(X\) to the departure set \(G_m=X-\bm e_m\Delta t\). The matrix \(P_m\) is sparse (each row contains only a few nonzero coefficients determined by a local stencil), but not unitary.

  This last fact is the obstruction that the present paper addresses. To embed the flow step into a quantum circuit one must block-encode the streaming matrix, and a generic sparse non-unitary matrix does not admit a cheap embedding: its block-encoding subnormalization $\alpha_{P_m}$ factor is of the order of the sparsity, and it requires \(\Theta(\log N)\) ancilla qubits \cite{lowchuang2019,gilyen2019}. Under the Carleman hierarchy these factors compound, degrading the success probability roughly as \(\alpha_{P_m}^{-N_C}\). We therefore seek a formulation in which the meshless translation is exactly unitary. The key observation is that a translation matrix can be obtained by evolving under a generator given by a linear combination of the derivative matrices $D_x$ and $D_y$. If \(D_x\) and \(D_y\) are skew-Hermitian, then the streaming operator is unitary. Therefore, it can be embedded with normalization factor \(\alpha=1\) and no ancilla, and can be time-stepped using standard quantum simulation machinery \cite{lloyd1996,lowchuang2019}. The difficulty is that RBF or least-squares derivative matrices built on a scattered set are not skew-Hermitian by construction.

  The key to restoring unitarity is to give the scattered node set a mimetic structure: following structure-preserving (mimetic) discretizations, which reproduce continuous operator identities such as integration by parts at the discrete level \cite{shashkov1996,lipnikov2014}, each node carries the area of its Voronoi cell---the discrete counterpart of the diagonal Hodge star in discrete exterior calculus \cite{desbrun2005}---and these masses define a discrete mass-weighted inner product. Under the corresponding similarity transformation, a derivative matrix that is skew-symmetric in this weighted inner product becomes skew-symmetric in the ordinary Euclidean inner product, so that its matrix exponential is an exactly unitary translation. The very same Voronoi weights determine how the distribution functions are encoded as quantum amplitudes, and we show that this encoding leaves the local collision rule intact up to an inconsequential rescaling of its quadratic term. For quasi-uniform node sets the Voronoi areas are mutually comparable, the mass-weighting diagonal then admits a block encoding with unit-order subnormalization and, crucially, a single ancilla. We validate the resulting mass-weighted meshless quantum LBM on the Taylor-Green vortex against the full meshless LBM, for Carleman truncation orders 1, 2 and 3, and demonstrate the expected improvement in accuracy as the truncation order increases.
  
  The closest quantum LBM constructions to our setting are the interpolation-supplemented QLBM of \cite{xiao2026} and its Taylor-series/least-squares fractional-step variant of \cite{yang2026}. Both, however, operate on a nonuniform Cartesian mesh rather than a genuinely scattered, meshless node set, and both replace the interpolation/derivative matrix by an SVD while assuming that the resulting dense unitary factors can be efficiently implemented, which is not justified for a general dense unitary matrix. Moreover, because a non-unitary collision operator must be reconstructed at every time step, the overall success probability compounds multiplicatively and decays exponentially in the number of steps. Our mass-weighted formulation avoids these bottlenecks: streaming operator is a single, exactly unitary that is reused at every time step, leaving only the local collision step to be embedded. Besides, Carleman linearization and the resulting history state avoid the formation of a new non-unitary collision operator at each time step.

  The remainder of the paper is organized as follows. Section~\ref{section 2} recalls the shifted incompressible LBM on uniform Cartesian lattices and its Carleman embedding.  In Section~\ref{section 3} we introduce the interpolation-supplemented meshless lattice Boltzmann method, expose the non-unitarity of its interpolation matrix, and explain the resulting difficulties for a quantum implementation. Section~\ref{section 5} develops the mimetic, mass-weighted derivative operators and the resulting unitary translation operator, together with the amplitude encoding of the distribution functions. The block-encoding cost is also analyzed. Numerical experiments on the Taylor-Green vortex are reported in Section~\ref{section 6}, and conclusions are drawn in Section~\ref{section 7}.

\section{Preliminaries}\label{section 2}
  \subsection{Lattice Boltzmann method}\label{2.1carlemanLBM}
    Let $\bm{f}(\bm{r},t)=\{ f_m(\bm{r},t)\}_{m=1}^{Q}$ as a vector function, and its component $f_m(\bm{r},t)$ represents the probability density of finding a fluid particle with a discrete velocity $\bm{e}_m$ at time $t$ and position $\bm{r}$. Local physical density $\rho$ and momentum $\rho \bm{u}$  are recovered as moments via
    \[
    \rho(\bm{r},t)=\sum_{m=1}^{Q}f_m(\bm{r},t),\qquad \rho \bm{u}(\bm{r},t)=\sum_{m=1}^{Q}f_m(\bm{r},t)\bm{e}_m.
    \]
    $f_m(\bm{r},t)$ satisfy the discrete Boltzmann equation (DBE)
    \[
    \frac{\partial f_m(\bm{r},t)}{\partial t} + \bm{e}_m\cdot \nabla f_m(\bm{r},t)= \Omega_m(\bm{r},t)
    \]
    where $\Omega_m(\bm{r},t)$ describes collision. The collision term can be given by the  Bhatnagar-Gross-Krook operator   
    \[
    \Omega_m(\bm{r},t)=-\frac{1}{\tau}[f_m(\bm{r},t)-f_m^{eq}(\bm{r},t)],
    \]
    where $\tau$ is the relaxation time. Here $f_m^{eq}(\bm{r},t)$ is the local Maxwell equilibrium expanded to second order
    \[
    f_{m}^{eq}(\bm{r},t)=\rho(\bm{r},t)\left(1+\frac{\bm{e}_m\cdot \bm{u}(\bm{r},t)}{c_s^2}+\frac{(\bm{e}_m\cdot \bm{u}(\bm{r},t))^2}{2c_s^4}-\frac{|\bm{u}(\bm{r},t)|^2}{2c_s^2}\right)\omega_m+\mathcal{O}(\text{Ma}^3),
    \]
    where $c_s$ is the speed of sound, $\text{Ma}=\frac{|\bm{u}|}{c_s}$ is the Mach number, and $w_m$ is the component of Maxwell weightings vector $\bm{w}$ satisfying $w_m\geq 0$ and $\sum_{m}w_m=1$.

    Integrating the discrete Boltzmann equation along its characteristics over one time step $\Delta t$, with the collision term discretized to second-order temporal accuracy, yields the lattice Boltzmann scheme
    \[
    \bar{f}_m(\bm{r}+\bm{e}_m\Delta t, t+\Delta t) = \bar{f}_m(\bm{r},t) - \frac{\Delta t}{\bar{\tau}}[\bar{f}_m(\bm{r},t)-f_m^{eq}(\bm{r},t)],
    \]
    where $\bar{f}_m(\bm{r},t)=f_m(\bm{r},t)-\frac{\Delta t}{2}\Omega_m(\bm{r},t)$, $\bar{\tau}=\tau+\frac{\Delta t}{2}$. The modified distribution function $\bar{f}_m(\bm{r},t)$ still satisfies the required moment conditions:
    \[
    \sum_{m=1}^{Q}\bar{f}_m=\sum_{m=1}^{Q}f_m=\rho \qquad \sum_{m=1}^{Q}\bar{f}_m \bm{e}_m=\sum_{m=1}^{Q}f_m \bm{e}_m=\rho \bm{u}.
    \] 
    In lattice units, the lattice spacing and time step are set to unity, i.e., $\Delta x=\Delta t=1$, resulting in the following lattice Boltzmann scheme:
    \begin{align}
    \bar{f}_m^{C}(\bm{r},t)&=\bar{f}_m(\bm{r},t)-\frac{1}{\bar{\tau}}[\bar{f}_m(\bm{r},t)-f^{eq}_m(\bm{r},t)],\label{eq1}
    \\
    \bar{f}_m(\bm{r}+\bm{e}_m,t+1)&=\bar{f}_m^{C}(\bm{r},t),\label{eq2}
    \end{align}
    where Eq.~\eqref{eq1} and \eqref{eq2} represent the collision and streaming processes, respectively.
    
    Furthermore, under the shifted incompressible assumption, we assume that $\rho=\rho_0+\delta \rho$ where $\rho_0=1$ is reference density and $\delta \rho$ denotes a small density fluctuation. We introduce the shifted distribution function $g_m(\bm{r},t)\coloneqq \bar{f}_m(\bm{r},t)-\omega_m$, which satisfies
    \[
    \sum_{m=1}^{Q}g_m=\delta\rho,\qquad 
    \sum_{m=1}^{Q}g_m\bm{e}_m=\rho\bm{u}. 
    \]
    The resulting lattice Boltzmann scheme can be written as
    \begin{align}
    g_m^C(\bm{r},t)&=g_m(\bm{r},t)-\frac{1}{\bar{\tau}}[g_m(\bm{r},t)-g_m^{eq}(\bm{r},t)],\\
    g_m(\bm{r}+\bm{e}_m,t+1) &= g_m^C(\bm{r},t).
    \end{align}
    Under the shifted incompressible assumption, macroscopic velocity is express as $\bm{u}= \sum_{m=1}^{Q}g_m\bm{e}_m$. The corresponding equilibrium distribution is
    \begin{align}
    g_m^{eq}(\bm{r},t)&=\bar{f}_m^{eq}(\bm{r},t)-\omega_m\approx\left[\delta\rho+3\bm{e}_m\cdot\bm{u}+\frac{9}{2}(\bm{e}_m\cdot\bm{u})^2-\frac{3}{2}|\bm{u}|^2\right]\omega_m \notag\\
    &=[\sum_{m_1=1}^{Q}(g_{m_1}+3 g_{m_1}E_{m,m_1}) + \sum_{m_1,m_2=1}^{Q}(\frac{9}{2}E_{m_1,m_1}E_{m,m_2}-\frac{3}{2}E_{m_1,m_2})g_{m_1}g_{m_2}]\omega_m,
    \end{align}
    where $c_s=1/\sqrt{3}$ has been used and $E_{m,n}:=\bm{e}_m\cdot \bm{e}_n$.

  \subsection{Carleman linearization}\label{Carleman linearization}
    The discrete distribution functions can be encoded into a quantum state.
    For a $D$-dimensional flow problem, let $N$ denote the total number of lattice nodes. The flow system at time $t$ can then be represented by a $d=NQ$-dimensional vector, which is encoded as the quantum state
    \begin{equation}
    \bm{g}(t) =\sum_{\bm{r}}\sum_{m=1}^{Q}g_m(\bm{r},t)|\bm{r}\rangle\otimes|m\rangle .\label{g(t)}
    \end{equation}
    Here, the first quantum register $|\bm{r}\rangle$ encodes the lattice position, while the second register $|m\rangle$ encodes the discrete velocity direction. In lattice units, the shifted incompressible lattice Boltzmann equation can be written as
    \begin{align}
   \bm{g}^{C}(t)
    &=(I+F_1)\bm{g}(t)+F_2\bm{g}(t)^{\otimes 2},\label{collision}\\
    \bm{g}(t+1)
    &=S\bm{g}^{C}(t),
    \end{align}
    where the first equation represents the collision step, while the second represents the streaming step. The definitions of $F_1$, $F_2$ and S are given below. 

    The collision step is the source of nonlinearity in the LBM and is spatially local, i.e., it acts independently at each lattice node and depends only on the discrete velocity directions. The corresponding operators $F_1$ and $F_2$ are given by
    \begin{align}
      F_1&=\sum_{\bm{r}}|\bm{r}\rangle\langle\bm{r}|\otimes\frac{1}{\bar{\tau}}\sum_{m,m_1=1}^{Q}\left[-\delta_{m,m_1}+\omega_m+3\omega_m E_{m,m_1}\right]|m\rangle\langle m_1|,
      \\F_2&=\sum_{\bm{r}}|\bm{r},\bm{r}\rangle\langle\bm{r}|\otimes\sum_{m,m_1,m_2=1}^{Q}\frac{\omega_m}{\bar{\tau}}\left[\frac{9}{2}E_{m,m_1}E_{m,m_2}-\frac{3}{2}E_{m_1,m_2}\right]|m\rangle\langle m_1,m_2|.
    \end{align}
    Since these operators are independent of the spatial position $\bm{r}$, they can be expressed as
    \[
    F_1=I_{\bm r}\otimes\check{F}_1,\qquad F_2=I_{\bm r,\bm r}\otimes\check{F}_2,
    \]
    where $I_{\bm r}\in\mathbb{R}^{N\times N}$ is the identity matrix and $I_{\bm r,\bm r}\in\mathbb{R}^{N\times N^2}$ satisfies
    \[
      I_{\bm r,\bm r}(i,j)= \{ \begin{array}{lll}
        1 & \text{if } & j=i^2 \\
        0 & \text{if } & \text{otherwise}
    \end{array}.
    \]
    It is worth noting that $\check{F}_1$ and $\check{F}_2$ are independent of time and are therefore fixed matrices.
    
    Meanwhile, the streaming operator $S$ transports the post-collision distribution functions to the corresponding neighboring lattice nodes and is given by
    \[
    S=\sum_{\bm{r}}\sum_{m=1}^{Q}|\bm{r}+\bm{e}_m\rangle\langle\bm{r}|\otimes|m\rangle\langle m|.
    \]
    The formulation above corresponds to the conventional lattice-based lattice Boltzmann method, in which the flow field is discretized on a regular lattice and the streaming operator transports the distribution functions between neighboring lattice nodes along the discrete velocity directions. In particular, the spatial structure of the streaming operator $S$ explicitly relies on the underlying regular lattice.

    The Carleman embedding introduces the lifted variables
    \[
    \bm{y}^{C,\infty}(t):=\left[\bm{g}^C(t),\bm{g}^C(t)^{\otimes 2},\bm{g}^C(t)^{\otimes 3},\ldots\right],\] 
    and 
    \[
    \bm{y}^{\infty}(t):=\left[\bm{g}(t),\bm{g}(t)^{\otimes 2},\bm{g}(t)^{\otimes 3},\ldots\right].
    \]
    
    \paragraph{Collision step}
    For the $k$th tensor power of the post-collision state, we have
    \begin{align}
      \bm{g}^C(t)^{\otimes k}&=\left[(I+F_1)\bm{g}(t)+F_2\bm{g}(t)^{\otimes2}\right]^{\otimes k}\nonumber\\&=\sum_{l=k}^{2k}\left[(I+F_1)^{\otimes(2k-l)}\otimes F_2^{\otimes(l-k)}+\mathrm{PERMS}\right]\bm{g}(t)^{\otimes l}\nonumber\\&:=\sum_{l=k}^{2k}C_l^k \bm{g}(t)^{\otimes l}.
    \end{align}
    Here, $\mathrm{PERMS}$ denotes the permutation terms arising from the different possible arrangements of the factors in the tensor product. In particular, it contains $\binom{k}{2k-l}$ terms, since tensor products are generally non-commutative with respect to the ordering of their factors. Therefore, the collision step can be written in matrix-vector form as
    \[
    \bm{y}^{C,\infty}(t)=\begin{bmatrix}
      \bm{g}^C(t)\\\bm{g}^C(t)^{\otimes2}\\\bm{g}^C(t)^{\otimes3}\\\vdots\end{bmatrix}
      =\begin{bmatrix}
        C_1^1&C_2^1&0&0&0&\cdots\\
        0&C_2^2&C_3^2&C_4^2&0&\cdots\\
        0&0&C_3^3&C_4^3&C_5^3&\cdots\\
        \vdots&\vdots&\vdots&\vdots&\vdots&\ddots
      \end{bmatrix}
      \begin{bmatrix}
        \bm{g}(t)\\\bm{g}(t)^{\otimes2}\\\bm{g}(t)^{\otimes3}\\\vdots
      \end{bmatrix}
      =:C^\infty \bm{y}^\infty(t),
    \]
    where 
    \[
    C_j^i\in\mathbb{R}^{N^i\times N^j},\qquad i\leq j\leq 2i.
    \]
    
    \paragraph{Streaming step}
    Since the streaming operator acts linearly on each tensor-product component,
    \[
    \bm{g}(t+1)^{\otimes k}=S^{\otimes k}\bm{g}^C(t)^{\otimes k}.
    \]
    Hence, the streaming step can be expressed in matrix-vector form as
    \[
    \bm{y}^\infty(t+1)=\begin{bmatrix}
      \bm{g}(t+1)\\\bm{g}(t+1)^{\otimes2}\\\bm{g}(t+1)^{\otimes3}\\ \vdots
    \end{bmatrix}=\begin{bmatrix}
      S&&&\\&S^{\otimes2}&&\\&&S^{\otimes3}&\\&&&\ddots
    \end{bmatrix}
    \begin{bmatrix}
      \bm{g}^C(t)\\\bm{g}^C(t)^{\otimes2}\\\bm{g}^C(t)^{\otimes3}\\ \vdots
    \end{bmatrix}
    =:S^\infty \bm{y}^{C,\infty}(t).
    \]
    Combining the collision and streaming steps, the shifted incompressible LBM can therefore be reformulated through the Carleman embedding as
    \[
    \bm{y}^\infty(t+1)=S^\infty C^\infty \bm{y}^\infty(t).
    \]
    Since $S^\infty$ and $C^\infty$ are infinite-dimensional matrices, we truncate the Carleman embedding at order $N_C$ by defining

     \[
    \bm{y}(t):=\left[\bm{g}(t),\bm{g}(t)^{\otimes2},\ldots,\bm{g}(t)^{\otimes N_C}\right].
    \]
    and
       \[
    \bm{y}^C(t):=\left[\bm{g}^C(t),\bm{g}^C(t)^{\otimes2},\ldots,\bm{g}^C(t)^{\otimes N_C}\right],
    \]
    The resulting truncated vector has dimension
    \[
    d+d^2+\cdots+d^{N_C}=\frac{d(d^{N_C}-1)}{d-1}.
    \]
    Correspondingly, truncating $S^\infty$ and $C^\infty$ to their first $N_C$ block rows and columns gives the finite-dimensional matrices $S$ and $C$. The resulting dynamics are approximated by
    \begin{equation}
    \bm{y}(t+1)\approx SC\bm{y}(t),\label{TCarleman}
    \end{equation}
    and hence
    \[
    \bm{y}(T)\approx(SC)^T \bm{y}(0).
    \]
    
    \paragraph{History state}
    To obtain the distribution function at a target time $T$ by directly solving a linear system, all intermediate time states can be collected into a single block linear system:
    \begin{equation}
    \begin{bmatrix}
      I&0&0&\cdots&0&0\\
      -SC&I&0&\cdots&0&0\\
      0&-SC&I&\cdots&0&0\\
      \vdots&\vdots&\vdots&\ddots&\vdots&\vdots\\
      0&0&0&\cdots&I&0\\
      0&0&0&\cdots&-SC&I
    \end{bmatrix}
    \begin{bmatrix}
      \bm{y}(0)\\\bm{y}(1)\\\bm{y}(2)\\\vdots\\\bm{y}(T-1)\\\bm{y}(T)
    \end{bmatrix}=
    \begin{bmatrix}
      \bm{y}(0)\\\bm{0}\\\bm{0}\\\vdots\\\bm{0}\\\bm{0}
    \end{bmatrix}.\label{HistoryLS1}
  \end{equation}
    
    By constructing a unitary embedding of the above linear system, a quantum linear system solver can be employed to obtain the history state containing the distribution functions at all time steps. The target state $\bm{y}(T)$ can then be extracted to recover the discrete distribution functions and, subsequently, the macroscopic flow field at time $T$.

\section{Quantum meshless LBM}\label{section 3}	
  \subsection{Interpolation-based meshless LBM}
    The conventional LBM typically relies on a regular lattice to discretize the computational domain, which can limit its applicability to problems involving complex geometries. To overcome this limitation, meshless lattice Boltzmann methods have been developed, in which the computational domain is represented by a set of scattered points rather than a regular lattice. 

    In a meshless spatial discretization, where the computational points may be nonuniformly distributed or randomly scattered, the streaming step transports the distribution function $g_m(\bm{r},t)$ at $\bm{r}$ to the position $\bm{r}+\bm{e}_m$. However, the resulting position $\bm{r}+\bm{e}_m$ generally does not coincide with any of the scattered points. Therefore, the conventional streaming operator needs to be modified. In contrast, the collision step is local in space and is therefore unaffected by the choice of spatial discretization.
    
    For a non-uniform scattered point set $X=\{\bm{x}_i\}_{i=1}^{N}$, we define a family of ghost point sets $\{G_m\}_{m=1}^{Q}$ by $G_m = X - \bm{e}_m$. Here, the scattered points are expressed in lattice units, with the lattice spacing taken as the reference length scale and $\Delta x=\Delta t=1$. In the meshless setting, $\Delta x$ does not represent the spacing between neighboring computational points; instead, $\bm{e}_m$ specifies the unit lattice displacement associated with the $m$th discrete velocity direction. An interpolation operator $\mathcal{P}_m: \mathbb{R}^{|X|}\to\mathbb{R}^{|G_m|}$ is constructed to interpolate the distribution function $g_m(t)$, originally defined on $X$, onto the ghost set $G_m$. The streaming step can then be rewritten as 
    \begin{equation}
    \bm{g}_m(X,t+1)
    = \bm{g}_m^C(X-\bm{e}_m,t)
    = \bm{g}_m^C(G_m,t)
    \approx P_m\,\bm{g}_m^C(X,t),
    \end{equation}
    where $P_m\in\mathbb{R}^{N\times N}$ is the interpolation matrix and the distribution function vectors are given by
    \begin{align}
    \bm{g}_m^C(X,t) &= \bigl(g_m^C(\bm{x}_1,t),\, g_m^C(\bm{x}_2,t),\,
    \ldots,\, g_m^C(\bm{x}_N,t)\bigr)^{T},\\
    \bm{g}_m(X,t) &= \bigl(g_m(\bm{x}_1,t),\, g_m(\bm{x}_2,t),\,
    \ldots,\, g_m(\bm{x}_N,t)\bigr)^{T}.
    \end{align}

    Under the shifted incompressible assumption as before, the streaming step of the meshless LBM can be expressed as 
    \begin{equation}
      \bm{g}(t+1) = \tilde{P}\bm{g}(t),
    \end{equation}
    where 
    \[
    \tilde{P} = \sum_{m=1}^{Q} P_m \otimes \lvert m\rangle\langle m\rvert \quad \text{and}\quad \bm{g}(t) =\sum_{i=1}^N\sum_{m=1}^{Q}g_m(\bm{x}_i,t)|\bm{x}_i\rangle\otimes|m\rangle .
    \]
    Compared with the streaming step in Eq.~\eqref{TCarleman}, the shift matrix $S$ is now replaced by an interpolation matrix $P$, where $P$ is obtained by truncating $P^{\infty}$ to its first $N_C$ block rows and block columns, with
    \begin{equation}
      P^{\infty} =
      \begin{bmatrix}
        \tilde{P} & & & \\
        & \tilde{P}^{\otimes 2} & & \\
        & & \tilde{P}^{\otimes 3} & \\
        & & & \ddots
      \end{bmatrix}.
    \end{equation}
     Furthermore,
    \begin{equation}
      \bm{y}(t+1) = P\,C\,\bm{y}(t).
    \end{equation}

    The corresponding linear system for solving the history state is modified as follows:
    \begin{equation}
    \begin{bmatrix}
      I&0&0&\cdots&0&0\\
      -PC&I&0&\cdots&0&0\\
      0&-PC&I&\cdots&0&0\\
      \vdots&\vdots&\vdots&\ddots&\vdots&\vdots\\
      0&0&0&\cdots&I&0\\
      0&0&0&\cdots&-PC&I
    \end{bmatrix}
    \begin{bmatrix}
      \bm{y}(0)\\\bm{y}(1)\\\bm{y}(2)\\\vdots\\\bm{y}(T-1)\\\bm{y}(T)
    \end{bmatrix}=
    \begin{bmatrix}
      \bm{y}(0)\\\bm{0}\\\bm{0}\\\vdots\\\bm{0}\\\bm{0}
    \end{bmatrix}.\label{HistoryLS2}
    \end{equation}

    There are various interpolation methods for scattered point sets, including RBF-local interpolation, moving least squares (MLS) and natural neighbor interpolation. A detailed discussion of these interpolation methods is provided in Section \ref{section 5}.
  
  \subsection{Data encoding}
  The Carleman vector is encoded in the following form:
  \[| y(t)\rangle=\frac{1}{\sqrt{N_C}}\sum_{s=1}^{N_C}
    | s\rangle\otimes | g(t)\rangle^{\otimes s}\otimes | 0\rangle^{N_C-s}.\]
  Correspondingly, the collision operator $C$ and streaming operator $S$ can be encoded as follows:
  \begin{align*}
  C&=\sum_{k=1}^{N_C}\sum_{l=k}^{\min\{2k,N_C\}}| k\rangle \langle l|
  \otimes C_{l}^{k}\otimes | 0\rangle^{\otimes l-k}\otimes I^{\otimes N_C-k}\\
  &:=\sum_{k=1}^{N_C}\sum_{l=k}^{\min\{2k,N_C\}}| k\rangle \langle l|
  \otimes \bar{C}_{l}^{k}\otimes I^{\otimes N_C-k},\\
  P&=\sum_{s=1}^{N_c}|s\rangle\langle s|\otimes \tilde{P}^{\otimes s}\otimes I^{\otimes N_C-s}.
  \end{align*}
  For simplicity, we denote the linear system in Eq.~\eqref{HistoryLS2} by
  \[
  A\bm{y}=\bm{b},
  \]
  where $A$ is encoded as
  \begin{equation}
  A = \sum_{t=0}^{T}|t\rangle\langle t|\otimes I -\sum_{0}^{T-1}|t+1\rangle \langle t|\otimes PC,\label{HisA}
  \end{equation}
  and $\bm{y}$ is encoded in a quantum state as 
  \[
  |y\rangle = \frac{1}{\sqrt{T+1}}\sum_{t=0}^{T}|t\rangle |y(t)\rangle.
  \]
  The number of qubits required to encode $|y\rangle$ is $q_A=\left\lceil\log(T+1)\right\rceil+\left\lceil\log N_C\right\rceil+N_C\left\lceil\log \right\rceil.$

  \subsection{Block-encoding of A}
    Our goal is to solve this linear system using a quantum linear system solver. Since $A$ is generally non-unitary, while quantum computers primarily operate on unitary operators, a unitary embedding of $A$, i.e., a block-encoding of $A$ is required, which is defined as a ($q_A+n_A$)-qubit unitary matrix $U_A$ such that
    \[
    \parallel A - \alpha_A(\langle 0|^{\otimes n_A} \otimes I  )U_A(| 0\rangle^{\otimes n_A} \otimes I )\parallel \leq \epsilon.
    \]
    Here, $U_A$ is referred to as an $(\alpha_A,n_A,\epsilon)$-block-encoding of matrix $A$.
    The complexity of the quantum linear system solver is closely tied to the parameters characterizing this unitary embedding.

    We next provide a brief discussion of the block-encoding parameters of  $A$. For simplicity, the approximation error is neglected by assuming $\epsilon=0$. The block-encoding parameters of $A$ can be analyzed by exploiting its recursive structure. To this end, we invoke the (linear combination of unitaries) LCU block-encoding lemma from \cite{Xiang2022}. Eq.~\eqref{HisA} can be further reformulated as
    \begin{equation}
    A=I-T_{-}\otimes (PC)
    \end{equation}
    where $T_{-}$ is the shift-down operator. There exists a $(1,1,0)$-block-encoding $U_{T_{-}}$ of $T_{-}$. Suppose that $U_{P}$ is an $(\alpha_{P},n_{P},0)$-block-encoding of $P$, and $U_C$ is an $(\alpha_C,n_C,0)$-block-encoding of $C$. Then, by the composition rule for block-encodings, there exists an $(\alpha_{P}\alpha_C,n_{P}+n_C,0)$-block-encoding $U_{PC}$ of $PC$. Furthermore, using the tensor-product rule for block-encodings, there exists an $(\alpha_{P}\alpha_C,n_{P}+n_C+1,0)$-block-encoding of $T_{-}\otimes PC$, where $T_{-}$ admits a $(1,1,0)$-block-encoding. Finally, by the LCU block-encoding lemma, there exists an 
    \begin{equation} 
      (\alpha_{P}\alpha_C+1,n_{P}+n_C+2,0)\label{blockencoding1}
    \end{equation}
    -block-encoding $U_{A}$ of $A=I-T_{-}\otimes PC$.

    The collision operator $C$ is identical to that of the lattice-based LBM\@. Therefore, based on the analysis of the block-encoding parameters of $C$ in~\cite{jennings2025a}, we have
    \begin{align}
      \alpha_C&=\sum_{\lfloor\frac{N_C}{2}\rfloor}^{l=0}\begin{pmatrix}N_C-l\\ l\end{pmatrix}\alpha_{I+F_1}^{N_C-2l}\alpha_{\bar{F_2}}^{l},\label{alphaC}\\
      n_C&=N_C-2+\log{(\lfloor\frac{N_C}{2}\rfloor+1)}+\max_{0\leq l\leq \lfloor \frac{N_C}{2}\rfloor}\left[\log{\begin{pmatrix}N_C-l\\ l\end{pmatrix}}+l(4D-1)\right],\label{nC}
    \end{align}
    where $\bar{F_2}=F_2\otimes |0\rangle$ is a square matrix. 
    In the lattice-based LBM, the streaming operator $S$ transports the distribution functions at each lattice node to the neighboring node along the corresponding discrete velocity direction. Therefore, under periodic boundary conditions, $S$ is unitary. In contrast, In contrast, interpolation matrices $P_m$ obtained from conventional interpolation methods for scattered point sets are generally non-unitary, and consequently, the resulting streaming operator $P$ is also non-unitary. We next analyze the block-encoding parameters of $P$.

    Suppose that, for each $m=1,\ldots,Q$, there exists an $(\alpha_{P_m},n_{P_m},0)$-block-encoding $U_{P_m}$ of the interpolation matrix $P_m$. Then, by $\tilde{P}=\sum_{m=1}^{Q}P_m\otimes |m\rangle \langle m|$, for block-encodings, there exists an $(\alpha_{\tilde{P}},n_{\tilde{P}},0)$-block-encoding of $\tilde{P}$, where
    \[
    \alpha_{\tilde{P}} =\max_{1\leq m\leq Q}\alpha_{P_m},\qquad n_{\tilde{P}}=\max_{1\leq m\leq Q}n_{P_m}+1.
    \]
    Furthermore, for each $i=1,\ldots,N_C$, there exists an $(\alpha_{\tilde{P}}^i,in_{\tilde{P}},0)$-block-encoding of $\tilde{P}^{\otimes i}$. Similarly, by $P=\sum_{s=1}^{N_c}|s\rangle\langle s|\otimes \tilde{P}^{\otimes s}\otimes I^{\otimes N_C-s}$, we can construct an $(\alpha_{P},n_{P},0)$-block-encoding of $P$, with
    \begin{align}
    \alpha_{P}&=\max_{1\leq i\leq N_C}\alpha_{\tilde{P}}^i=(\max_{1\leq m\leq q}\alpha_{P_m})^{N_C},\label{alphaP}\\
    n_{P}&=\max_{1\leq i\leq N_C}in_{\tilde{P}}+1=N_C\max_{1\leq m\leq q}n_{P_m}+N_C+1.\label{nP}
    \end{align}
  
    Obviously, due to the non-unitary streaming operate $P$, compared to the block-encoding parameters $(\alpha_C+1,n_C+2,0)$ of the coefficient matrix obtained by the regular lattice-based LBM, the block-encoding of the meshless LBM requires more quantum resources. In particular, the success probability is reduced to the order of \(1/\alpha_P^2\), where \(\alpha_P\) grows exponentially with the Carleman truncation order. Meanwhile, the number of required ancillary qubits increases by \(n_P\), where \(n_P\) grows linearly with the Carleman truncation order. The resulting quantum linear-system complexity described by the number of queries to $U_A$, $U_A^\dagger$ and their controlled versions, depends on both the normalization factor of the block-encoding, the condition number $\kappa_A$ of the resulting linear system and the related error $\epsilon_Q$ required\cite{Dalzell2024}
    \[
    q_Q\leq \frac{\alpha_A}{\parallel A \parallel}\left(56\kappa_A+1.05\kappa_A\ln{\frac{\sqrt{1-\epsilon_Q^2}}{\epsilon_Q}}+2.78\ln{\kappa_A}^3+3.17\right).
    \]

    Therefore, to reduce the complexity of the quantum linear-system solver, a natural question is whether the properties or structure of the interpolation operators \(P_m\) can be modified to reduce the overhead associated with their block-encodings. An even more desirable possibility is to construct unitary interpolation operators, thereby eliminating the need for block-encoding altogether.

\section{Unitarization of the translation interpolation matrices}\label{section 5}

  First, it is observed that the interpolation matrix $P_m$ does not represent an interpolation from the set of data points $X$ to any other random set of points. The set of evaluation points $G_m=X-\bm{e}_m$ is a global translation of the set of data points $X$. In other words, the interpolation matrix $P_m$ can be viewed as the finite-dimensional restriction of the continuous translation operator $\mathcal{P}_m$, defined by 
  \[
  (\mathcal{P}_m f)(\bm{x})=f(\bm{x}-\bm{e}_m).\]
  And under periodic boundary conditions, the translation operator $\mathcal{P}_m$ is unitary, which suggests that it is reasonable to seek a unitary interpolation operator $P_m$. Furthermore, from the Taylor expansion of a multivariate function:
  \[
  f(\bm{x}+\bm{e})=\exp(\bm{e}\cdot\nabla)f(\bm{x})=\exp(\sum_{i=1}^{D}e_i\nabla_i)f(\bm{x})
  \]
  where $\nabla_i$ denotes the derivative operator along the $i$-th spatial direction. Therefore, letting $D = 2$ for simplicity and defining the symbol $\bm{e}_m=(a,b)$, we can construct the interpolation operator using
  \[
  P_m=\exp(-(aD_x+bD_y))
  \]
  where $D_x$ and $D_y$ are the discrete derivative operators along the $x$ and $y$ spatial directions, respectively. 
  We therefore seek a discrete unitary operator that approximates the same flow map, and we obtain it by discretizing the generator rather than by projecting the interpolation matrix onto the unitary group.

  \subsection{The Voronoi mass matrix as the canonical metric}
    For non-uniform scattered points, the standard equally weighted inner product is generally inappropriate. Instead, a weighted inner product should be employed to account for the non-uniform distribution of the points.
    For an exponential \(\exp(A)\) to be unitary, the generator \(A\) must be skew-Hermitian with respect to the inner product in which unitarity is measured. This condition is not intrinsic to the matrix \(A\): if the space of nodal functions is equipped with the inner product \(\langle \bm{g},\bm{f}\rangle_G=\bm{g}^\dagger G \bm{f}\) for a positive definite \(G\), then \(A\) is skew-Hermitian if and only if
    \[
    G A + A^\dagger G = 0 .
    \]
    If the Euclidean metric $G = I$ is chosen by default, this skew-Hermitian matrix satisfies $A^\dagger=-A$, which is consistent only with a uniform lattice. On scattered nodes is related to the distribution of the scatter points.

    Voronoi-cell volumes, $\{m_i\}_{i=1}^N$, can be used as quadrature weights to account for the non-uniform scattered points $X=\{\bm{x}_i\}_{i=1}^N\subseteq\mathbb{R}^D$, leading to a more appropriate discrete inner product. They are positive, partition the domain, \(\sum_i m_i=|\Omega|\) where $|\Omega|$ is the total volume of the problem domain, and reduce to the uniform measure \(m_i\equiv h^D\) on a regular lattice, where $h$ is the lattice spacing. This makes
    \[
    M=\operatorname{diag}(m_1,\dots,m_N),\qquad
    \langle \bm{g},\bm{f}\rangle_M=\sum_i m_i\,\bar{g_i}f_i
    \]
    the unique discrete inner product that is consistent with the continuum \(L^2(\Omega)\) inner product as \(h\to0\). We therefore require the discrete generators to be skew-Hermitian in this metric.

  \subsection{Mimetic skew-Hermitian discretization of the generators.}
    For each node $\bm{x}_i$, we define its stencil as the set of its $q$ nearest neighbors under periodic boundary conditions and Euclidean distance, excluding the point itself. The collection of all such stencils induces a directed graph $G=(X,E)$, where a directed edge $e=(i,j)\in E$ if and only if $\bm{x}_j$ belongs to the stencil of $\bm{x}_i$. To obtain a symmetric stencil structure, we augment the edge set by enforcing reciprocity: for every directed edge $e=(i,j)\in E$, the reverse edge $(j,i)$ is also included in $E$; that is, if $\bm{x}_j$ is in the stencil of $\bm{x}_i$, then $\bm{x}_i$ is also in the stencil of $\bm{x}_j$. Therefore, the new stencil size  of each node is between $q$ and $2q$. For each point \(\bm{x}_i\), we define its stencil point index set \(\mathrm{Ste}(i)\), where \(j\in\mathrm{Ste}(i)\) if and only if \((i,j)\in E\).
    The total number of directed edges in the resulting symmetric graph is denoted by $N_E=|E|$.  Below is a step-by-step process for obtaining skew-Hermitian generators:


    \textbf{Step 1} — skew-Hermitianization of the discrete generators.  For every undirected edge \(e=(i,j)\), \(i<j\), there is a scalar weight \(w_{ij}\in\mathbb R\). And the generators \(D_x,D_y\in\mathbb R^{N\times N}\) are assembled by
    \[
    [D]_{ij}=\frac{w_{ij}}{m_i},\qquad [D]_{ji}=-\frac{w_{ij}}{m_j},\qquad [D]_{kk}=0 .
    \] 
    By construction the row-scaled matrices are exactly skew-symmetric,
    \[
    MD+D^{\top}M=0 ,
    \]
    so each \(D\) is skew-symmetric in the mass-weighted inner product \(\langle \bm{g},\bm{f}\rangle_M=\sum_i m_i g_i f_i\). This parameterization guarantees that the skew-symmetry holds identically in the weights $\{w_{ij}\}$, which are the only free parameters left to optimize. 

    \textbf{Step 2} — weight optimization. Let $D$ be the discrete matrix obtained by taking the derivative along the direction $d$ where $d$ is $(1,0)$ or $(0,1)$. First, we require the weights to exactly reproduce the derivatives of constant and linear functions, which leads to the following moment conditions:
    \begin{align*}
    \sum_{j\in\mathrm{Ste}(i)}w_{ij}&=0,\\
    \sum_{j\in\mathrm{Ste}(i)} w_{ij}\,(\boldsymbol x_j-&\boldsymbol x_i)=m_i\,d ,
    \end{align*}
    where it is worth noting that $\boldsymbol x_j-\boldsymbol x_i$ is subtracted under periodic boundary conditions. These rows make \(D\,\mathbf 1=0\) and \(D(\boldsymbol d_0\cdot\boldsymbol x)\equiv \boldsymbol d_0\cdot d\) for any constant vector \(\boldsymbol d_0\);
    In the implementation the unknowns are stored once per undirected edge, \(e=\{i,j\}\) with \(i<j\), and the oriented value in row \(i\) is recovered as \(w_{ij}=\operatorname{sgn}(j-i)w_{\min(i,j),\max(i,j)}\); this identity is equivalent to \(w_{ji}=-w_{ij}\).

    Meanwhile, we require the weights to approximately reproduce the derivatives of low-frequency Fourier modes, leading to the following moment conditions:
    \begin{align*}
    [D\cos(2\pi \boldsymbol p\cdot\boldsymbol x)](\boldsymbol x_i)\simeq
    \partial_d\cos(2\pi\boldsymbol p\cdot\boldsymbol x)\big|_{\boldsymbol x_i},\\    
    [D\sin(2\pi \boldsymbol p\cdot\boldsymbol x)](\boldsymbol x_i)\simeq
    \partial_d\sin(2\pi\boldsymbol p\cdot\boldsymbol x)\big|_{\boldsymbol x_i},
    \end{align*}
    for \(\boldsymbol p\in\{-2,\dots,2\}^2\setminus\{0\}^2\). 

    All of the above constraints can be formulated as a least-squares problem:
    \[
    \min_{w\in\mathbb{R}^{N_E/2}}\ \xi^2\bigl\|A_{\mathrm{mom}}w-b_{\mathrm{mom}}\bigr\|^2+\bigl\|A_{\mathrm{mod}}w-b_{\mathrm{mod}}\bigr\|^2+\gamma\,\|w\|^2,
    \]
    where $\xi \gg 1$. The moment conditions are strongly enforced by assigning a large penalty weight \(\xi^2\), while the Fourier-mode reproduction conditions are imposed as soft constraints. All rows are normalized before assembly, and the ridge term \(\gamma\|w\|^2\)
    controls the spectral norm of the generator, i.e. the norm of the quantum Hamiltonian.

    \textbf{Step 3} — encoded (Euclidean) anti-symmetric generators. Define
    \[
    \widetilde D=M^{1/2}DM^{-1/2}.
    \]
    Because \(MD+D^{\top}M=0\), each \(\widetilde D\) is exactly skew-symmetric in the Euclidean metric, i.e. \(\widetilde D^{\top}=-\widetilde D\).

    Note that \(P_m = \exp(-(aD_x+bD_y))\) is not Euclidean-unitary. Therefore, the operator executed on the quantum register is its conjugate
    \[
    \widetilde P_m=M^{1/2}P_mM^{-1/2}
    =\exp(-(a\,\widetilde D_x+b\,\widetilde D_y)),
    \qquad \widetilde D_j=M^{1/2}D_jM^{-1/2},
    \]
    where each \(\widetilde D_j\) is real skew-symmetric and hence \(-i\widetilde D_j\) is Hermitian; equivalently \(\widetilde P_m=e^{-iH}\) with Hermitian \(H = -i(a\,\widetilde D_x+b\,\widetilde D_y)\).  Moreover, $\widetilde P_m$ can be implemented efficiently by quantum Hamiltonian simulation. For instance, qubitization block-encodes \(H\) and converts it into the unitary \(e^{-iH}\) through quantum singular value transformation (QSVT) \cite{lowchuang2019, gilyen2019}, while a  Trotter-Suzuki product \(e^{aD_x\Delta t/r}e^{bD_y\Delta t/r}\), raised to the \(r\)-th power, offers a direct splitting of the two one-dimensional skew terms \cite{suzuki1990}.
    
    \subsection{The new amplitude encoding with mass weight}
    Accordingly, the same conjugation fixes the amplitude encoding. Recalculate the vector of probability distribution functions via amplitude encoding:
    \[
    \tilde{\bm{ g}}(t)= (\sum_m M^{1/2}\otimes |m\rangle\langle m|) \bm{g}(t)=\sum_m\sum_{\bm{r}}\sqrt{m_{\bm{r}}}g(\bm{r},\bm{e}_m)|\bm{r}\rangle |m\rangle,
    \]
    where the definition of $\bm{g}(t)$ is given in Eq.~\eqref{g(t)}. Similarly, re-encode post-collision distribution functions vector as:
    \[
     \tilde{\bm{g}}^C(t)= (\sum_m M^{1/2}\otimes |m\rangle\langle m|) \bm{g}^C(t)=\sum_m\sum_{\bm{r}}\sqrt{m_{\bm{r}}}g^C(\bm{r},\bm{e}_m)|\bm{r}\rangle |m\rangle.
    \]
    We formulate the collision process between the mass-weighted pre-collision probability density distribution function $\bm{\tilde g}$ and thepost-collision probability density distribution function $\bm{\tilde g}^C(t)$ under the shifted incompressible assumption. From $\tilde g_m=\sqrt{m_{\bm{r}}}\,g_m$ the re-encoded density fluctuation and reference density are
    \[
    \delta\tilde\rho=\sum_m\tilde g_m=\sqrt{m_{\bm{r}}}\,\delta\rho,
    \qquad
    \tilde\rho_0=\sqrt{m_{\bm{r}}}\,\rho_0,
    \]
    while the re-encoded velocity,
    \[
    \tilde{\boldsymbol u}=\frac{\sum_m\boldsymbol e_m\tilde g_m}{\tilde\rho}
    =\frac{\sqrt{m_{\bm{r}}}\sum_m\boldsymbol e_m g_m}{\sqrt{m_{\bm{r}}}\,\rho}
    =\boldsymbol u,
    \]
    is unchanged. However, under the shifted incompressible assumption, the re-encoded velocity is recovered as 
    \[
    \tilde{\boldsymbol u}=\sum_m\boldsymbol e_m\tilde g_m
    =\sqrt{m_{\bm{r}}}\boldsymbol u.
    \]
    Substituting these into the collision step formula, we have
    \begin{align*}
        \tilde g_m^C(\bm{r},t)&= \sqrt{m_{\bm{r}}} g^C_m=\sqrt{m_{\bm{r}}}(g_m-\frac{1}{\bar{\tau}}[g_m-g_m^{eq}])=\tilde{g}_m-\frac{1}{\bar{\tau}}[\tilde{g}_m-\sqrt{m_{\bm{r}}}g_m^{eq}]\\
        &=\tilde{g}_m-\frac{1}{\bar{\tau}}[\tilde{g}_m-\omega_m(\delta \tilde{\rho}+3\bm{e}_m\cdot \tilde{\boldsymbol u}+\frac92\frac{1}{\sqrt{m_{\bm{r}}}}(\bm{e}_m\cdot \tilde{\boldsymbol u})^2-\frac32\frac{1}{\sqrt{m_{\bm{r}}}}|\tilde{\boldsymbol{u}}|^2)].
    \end{align*}
    As can be seen, compared to the classic LBM format, the quadratic term in the mass-weighted meshless LBM collision operator has a coefficient of $\frac{1}{\sqrt{m_{\bm{r}}}}$, which is related to the mass weight of that point. For a regular lattice, this coefficient is the same at every point, so the formulation degenerates to the classical LBM. 

    Therefore, after mass-weighted encoding of the distribution function values at the scattered points, the resulting encoded distribution functions satisfy the rewrote collision formula:
    \begin{equation}
    \tilde{\bm{g}}^{C}(t) =(I+F_1)\tilde{\bm{g}}(t)+\tilde{F}_2 \tilde{\bm{g}}(t)^{\otimes 2}.\label{mass weighted collision}
    \end{equation}
    Here, $\tilde{F}_2=(M^{-1/2}\otimes I)F_2$, while $F_1$ and $F_2$ are defined as in Eq.~\eqref{collision}.

    In addition, under this amplitude encoding, the interpolation matrices still form the streaming matrix via $\tilde{P} = \sum_{m=1}^{Q} \widetilde P_m \otimes \lvert m\rangle\langle m\rvert $. We have
    \begin{equation}
    \begin{aligned}
    \tilde P \tilde{\bm{g}}^C(t) &= (M^{1/2}\otimes I) (\sum_{m}  P_m \otimes \lvert m\rangle \langle m\rvert) (M^{-1/2}\otimes I) (M^{1/2}\otimes I) \bm{g}^C(t)\\
    &=(M^{1/2}\otimes I) (\sum_{m} P_m \otimes \lvert m\rangle \langle m\rvert)  \bm{g}^C(t)\\
    &=(M^{1/2}\otimes I) \bm{g}(t+1)=\tilde{\bm{g}}(t+1),
    \end{aligned}\label{mass weighted streaming}
  \end{equation}
    where $\widetilde P_m$ is Euclidean-unitary. 
    The unitary streaming step in Eq.~\eqref{mass weighted streaming}, combined with the collision step in Eq.~\eqref{mass weighted collision}, defines the mass-weighted meshless lattice Boltzmann method (MW-MLBM) proposed in this work.
    
    Therefore, the linear system for the re-encoded historical state is
    
    \begin{equation}
    \tilde{A}\tilde{\bm{y}}=\begin{bmatrix}
      I&0&0&\cdots&0&0\\
      -\tilde{P}\tilde{C}&I&0&\cdots&0&0\\
      0&-\tilde{P}\tilde{C}&I&\cdots&0&0\\
      \vdots&\vdots&\vdots&\ddots&\vdots&\vdots\\
      0&0&0&\cdots&I&0\\
      0&0&0&\cdots&-\tilde{P}\tilde{C}&I
    \end{bmatrix}
    \begin{bmatrix}
      \tilde{\bm{y}}(0)\\\tilde{\bm{y}}(1)\\\tilde{\bm{y}}(2)\\\vdots\\\tilde{\bm{y}}(T-1)\\\tilde{\bm{y}}(T)
    \end{bmatrix}=
    \begin{bmatrix}
      \tilde{\bm{y}}(0)\\\bm{0}\\\bm 0\\\vdots\\\bm 0\\\bm 0
    \end{bmatrix}=\tilde{\bm b}.\label{HistoryLS3}
    \end{equation}
    where $\tilde{C}$ is the collision operator obtained by Carleman linearization in the same manner as described in Section \ref{Carleman linearization}, and $\tilde{\bm y}(t):=\left[\tilde{\bm g}(t),\tilde{\bm g}(t)^{\otimes2},\ldots,\tilde{\bm g}(t)^{\otimes N_C}\right]$. Furthermore, the block encoding of the coefficient matrix $\tilde{A}$ is consistent with that in Section \ref{section 3}, yielding the $(\alpha_{\tilde A}, n_{\tilde A}, 0)$ block encoding $U_{\tilde A}$ of matrix $\tilde A$ with
    \[
    \alpha_{\tilde A} = \alpha_{\tilde{C}} +1,\qquad n_{\tilde A} = n_{\tilde{C}}+2.
    \]
    Since the only difference between $C$ and $\tilde{C}$ lies in $F_2$ and $\tilde{F}_2$ during the construction process, we can see from Eq.~\eqref{alphaC} and \eqref{nC} that
    \begin{align}
      \alpha_{\tilde{C}}&=\sum_{\lfloor\frac{N_C}{2}\rfloor}^{l=0}\begin{pmatrix}N_C-l\\ l\end{pmatrix}\alpha_{I+F_1}^{N_C-2l}\alpha_{\hat{F}_2}^{l},\label{alphaC'}\\
      n_{\tilde{C}}&=N_C-2+\log{(\lfloor\frac{N_C}{2}\rfloor+1)}+\max_{0\leq l \leq \lfloor \frac{N_C}{2}\rfloor}\left[\log{\begin{pmatrix}N_C-l\\ l\end{pmatrix}}+(N_C-2l)n_{I+F_1}+ln_{\hat{F}_2}\right],\label{nC'}
    \end{align}
    where $\hat{F}_2=\tilde{F}_2\otimes|0\rangle=(M^{-1/2}\otimes I\otimes I)\bar{F}_2$. Suppose $M^{-1/2}$ has an $(\alpha_M,n_M,0)$-block-encoding $U_M$; then there exists an $(\alpha_{\hat{F}_2},n_{\hat{F}_2},0)$-block-encoding $U_{\hat{F}_2}$ of $\hat{F}_2$ that satisfies 
    \[
    \alpha_{\hat{F}_2} = \alpha_M\alpha_{\bar{F}_2},\qquad n_{\hat{F}_2}=n_{\bar{F}_2}+n_M.
    \]
    Therefore, the block-encoding parameters of $\tilde C$ and $C$ are related as $\alpha_{\tilde{C}}\leq \alpha_M^{\lfloor\frac{N_C}{2}\rfloor}\alpha_C, n_{\tilde{C}}\leq n_C+\lfloor\frac{N_C}{2}\rfloor n_M$. It can be seen that the success probability is reduced to the order of \(1/\alpha_M^{N_C}\), which decreases exponentially with the Carleman truncation order. Fortunately, the parameter $\alpha_M$ can be kept at the $\mathcal{O}(1)$ level by controlling the degree of scatteredness of the scatter points. Note that the absolute magnitude of the mass weights $m_i$ carries no physical content: only their ratios do. Therefore, by globally rescaling the weights $m_i$ and redefining them as
    \begin{align*}
    m_i \leftarrow \frac{m_i}{\text{mean}_{1\leq i\leq N}\{m_i\}},
    \end{align*}
    all the weights remain of order $\mathcal{O}(1)$ for quasi-uniformly distributed scattered points. And by the one-ancilla unitary dilation, the block-encoding parameters of diagonal matrix $M^{-1/2}$ satisfy $\alpha_M=\parallel M^{-1/2}\parallel_{\infty}=\max_{i} \frac{1}{\sqrt{m_i}}\sim\mathcal{O}(1)$ and $n_M=1$. 
    For contrast, consider block-encoding the raw, non-unitary interpolation matrix \(P_m\) before mass weighting. Although each row of \(P_m\) carries at most \(k\) nonzero entries, sparsity alone does not produce a cheap embedding. In the sparse-access or LCU construction, a generic \(s\)-sparse matrix with unit maximum entry requires a subnormalization factor \(\alpha=\Theta(s)\), so here \(\alpha_{P_m}=\Theta(k)\); in particular there is no structural reason for \(\alpha_{P_m}=\mathcal{O}(1)\), since \(\alpha_{P_m}\) must cover \(\|P_m\|\) and \(P_m\) is not unitary. The same constructions also demand \(\Theta(\log N)\) index qubits, and the single-ancilla dilation is available only for diagonal operators, so a general non-unitary sparse matrix cannot be embedded exactly with one ancilla.

    In short, the mass weighting restores unitarity of the flow step and thereby collapses its block-encoding overhead from \(\Theta(k)^{N_C}\) normalization and \(N_C \Theta(\log N)\) ancillas to \(\mathcal{O}(1)^{\lfloor\frac{N_C}{2}\rfloor}\) and $\lfloor\frac{N_C}{2}\rfloor$ ancillas.

\section{Numerics}\label{section 6}
  \subsection{Accuracy of the unitary translation interpolation matrix}
    We first study the interpolation accuracy of the unitarized translation interpolation matrix through numerical experiments. We test the accuracy of the unitary interpolation matrix on the two-dimensional square domain \([0,1]\times[0,1]\). Two types of scattered point sets are considered in this study. The first is generated by perturbing a regular grid. The grid spacing \(h\) is used to characterize the density of the scattered points, and the perturbation is given by $\delta h\,\mathrm{rand},$ where \(\mathrm{rand}\) is a uniformly distributed random number in \([-1,1]\). While the second consists of Halton points, which are more irregularly distributed than the perturbed regular grid. The test function is
    $
    f(x,y)=\sin(2\pi x)\cos(2\pi y),
    $
    which is periodic over the computational domain. The evaluation point set is defined as $G_m=X+\bm{e}_m$, where $\bm{e}_m=(h,h)$. Define \(\bm{f}(X)\) and \(\bm{f}(G_m)\) as the vectors of function values of \(f\) evaluated on \(X\) and \(G_m\) respectively, while \(\tilde{\bm{f}}(X)\) and \(\tilde{\bm{f}}(G_m)\) the corresponding vectors of mass-weighted function values. The interpolation error is defined as
    $$
    \mathrm{err}=\frac{\left\|P_m \tilde{\bm{f}}(X)-\tilde{\bm{f}}(G_m)\right\|_2}{\left\|\tilde{\bm{f}}(G_m)\right\|_2},
    $$
    where \(P_m\) denotes the unitary interpolation matrix from the point set $X$ to $G_m$ obtained from the unitarization of interpolation described above. For comparison, we evaluate the accuracy of the RBF-local interpolation under periodic boundary conditions, where the interpolation is performed directly on the original function-value vector \(\bm{f}(X)\), without mass weighting. We also evaluate the accuracy  obtained by directly using the streaming operator \(S\) from the lattice-based LBM. Since the scattered points considered here are not completely random but are generated by perturbing a regular grid, the streaming operator \(S\) transports the function values on \(X\) to points near the corresponding points in \(G_m\). The experimental results are shown in Table \ref{tab:unitarity}
    \begin{table}[htbp]
    \centering
    \footnotesize
      \begin{tabular}{c c c c c c c}
        \toprule
        Method & $\delta h=0$ & $\delta h=0.1h$ & $\delta h=0.2h$ & $\delta h=0.3h$ & Halton & Unitary \\
        \midrule
        Mass Weighted & 2.17e-05 & 2.93e-04& 4.90e-04& 6.46e-04& 9.87e-04 & Yes\\
        RBF-local & 1.19e-04 & 1.15e-04 &1.13e-04 & 1.16e-04 &1.13e-04& No \\
        Natural Neighbor & 4.44e-16 & 1.42e-02 & 2.85e-02 & 4.27e-02 &1.66&Yes \\
        \bottomrule
      \end{tabular}
    \caption{A comparison of the accuracy of translation interpolation using different interpolation methods on scatter sets with varying degrees of irregularity. In these experiments, the number of data points is 2500, corresponding to $h = 1/50$. For both the mass-weighted unitary interpolation method and the RBF method, the stencil size is $q = 20$.}
    \label{tab:unitarity}
    \end{table}
    As can be seen, compared to RBF interpolation, our mass-weighted interpolation method preserves unitary properties and exhibits a gradual decline in accuracy as the degree of irregularity increases. In contrast, the accuracy of RBF interpolation is insensitive to the degree of irregularity of the point set. However, the overall accuracy of our mass-weighted interpolation method remains on the same order of magnitude as that of RBF interpolation. Meanwhile, using the streaming operator from the regular-grid LBM method as the unitary interpolation matrix yields exact results on regular grids but suffers from a drastic deterioration in accuracy as the degree of irregularity increases. And the accuracy is completely lost on the Halton point set.

  \subsection{Accuracy of the full MW-MLBM}\label{sec:tgv-feasibility}
    Regarding our proposed MW-MLBM method, we first test the accuracy of the full MW-MLBM without the Carleman linearization on MATLAB, which follows the collision and streaming step described in Eq.~\eqref{mass weighted collision} and \eqref{mass weighted streaming}. We simulate the Taylor-Green (TG) flow with the MW-MLBM and, for reference, with the regular-lattice LBM. Both methods share the same shifted incompressible, quadratic collision model introduced above and differ only in the spatial discretization and in the streaming step. The meshless method places the distribution function on a perturbed point cloud \(X=\{\boldsymbol x_i\}_{i=1}^{N}\), endows each node with its Voronoi mass \(m_i\), and yields the unitary translation operators \(S_q=\exp( a\widetilde D_x+ b\widetilde D_y)\) for the eight moving directions. The lattice baseline uses exact permutation streaming on the regular \(n\times n\) grid. All physical parameters are identical for the two schemes.

    We consider the decaying TG vortex in the periodic domain \([0,1]\times [0,1]\) with initial velocity field
    \begin{equation}
    \boldsymbol u_0 = U_0\bigl(\sin(2\pi x)\cos(2\pi y),\,
    -\cos(2\pi x)\sin(2\pi y)\bigr),
    \end{equation}
    whose incompressible Navier-Stokes solution and kinetic energy read

    \begin{equation}
    \boldsymbol u^{th}(\boldsymbol x,t)=\boldsymbol u_0(\boldsymbol x)\,e^{-8\pi^2\nu t},
    \qquad
    E^{th}(t)=\frac{U_0^2}{4}\,e^{-16\pi^2\nu t}.
    \end{equation}
    The velocity scale and Reynolds number are fixed at \(U_0=0.05\) (\(\mathrm{Ma}=U_0/c_s\approx0.087\)) and \(\mathrm{Re}=U_0L/\nu=50\) with \(L=1\), so that the kinematic viscosity \(\nu=U_0/\mathrm{Re}\) is independent of the resolution. For each resolution \(n\) with \(h=1/n\) and \(N=n^2\) points, the relaxation parameter is set accordingly, \(\bar{\tau}=\tfrac12+3\nu/h\). The meshless clouds are jittered lattices ($\delta h =0.1 h$) with \(k=12\) nearest neighbors, using a single realization per \(n\) with fixed random seed. Each simulation advances one advection time, \(T_{\mathrm{end}}=L/U_0=20\), i.e.\ \(N_t=\lceil T_{\mathrm{end}}/h\rceil\) steps. Errors are measured in the mass-weighted relative \(L_2\) norm
    \begin{equation}
    \varepsilon_{\mathrm{rel}}(T_{\mathrm{end}}) =\frac{\sqrt{\sum_i m_i\bigl|\boldsymbol u_i(T_{\mathrm{end}}) -\boldsymbol u_i^{\mathrm{th}}(T_{\mathrm{end}})\bigr|^2}} {\sqrt{\sum_i m_i\bigl|\boldsymbol u_i^{\mathrm{th}}(T_{\mathrm{end}})
    \bigr|^2}}, \label{L2err}
    \end{equation}
    while the kinetic energy is evaluated by the Voronoi quadrature 
    \[
    E=\tfrac12\sum_i m_i\rho_i|\boldsymbol u_i|^2/\sum_i m_i, 
    \]
    and the measured viscosity \(\nu_{\mathrm{meas}}\) is extracted from a linear fit of \(\ln E(t)\).

\begin{figure}[htbp]
      \centering
      \includegraphics[width=1\linewidth]{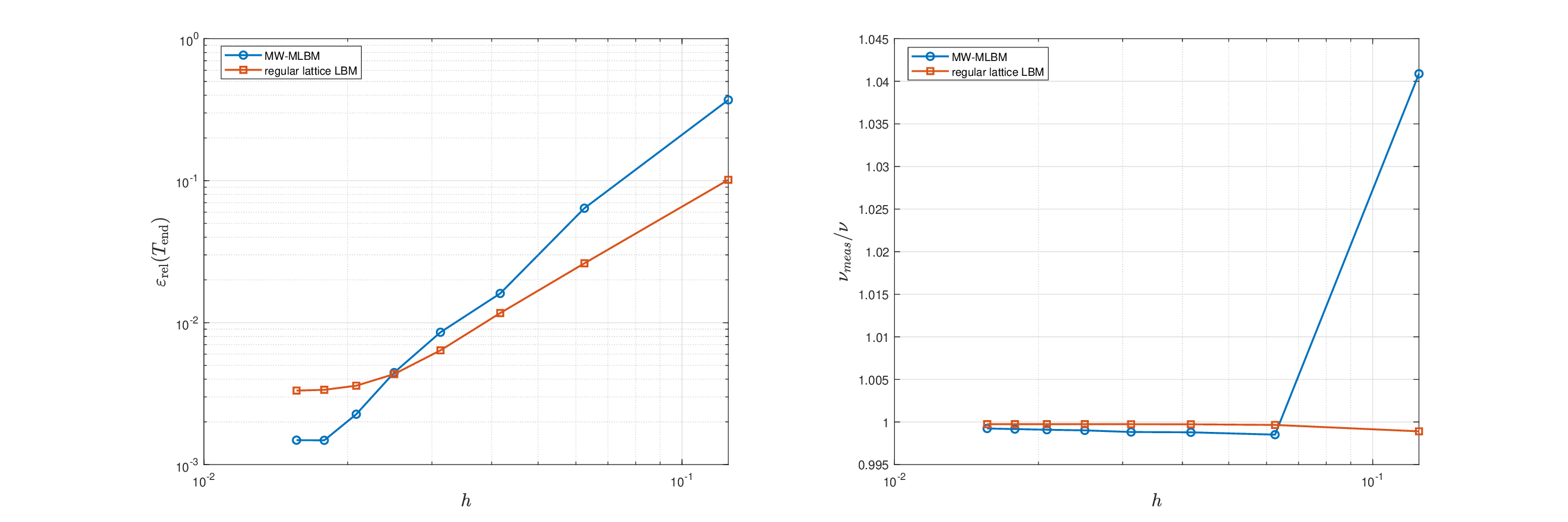}
      \caption{Convergence of the decaying TG simulation at fixed
      \(\mathrm{Re}=50\), \(U_0=0.05\), after one advection time.
      Left: mass-weighted relative \(L_2\) velocity error at the final time versus
      \(h\), with fitted slopes of about \(2.7\) (meshless, \(n\ge24\)) and
      \(1.9\) (regular lattice).
      Right: fitted viscosity ratio \(\nu_{\mathrm{meas}}/\nu\)
      versus \(h\).}
      \label{fig:tgv-conv}
    \end{figure}

    The numerical results are collected in Fig.~\ref{fig:tgv-conv}. The errors of both schemes decrease with \(h\). The lattice baseline converges as \(\varepsilon_{\mathrm{rel}}\sim O(h^{1.9})\), consistent with the expected second-order accuracy of the LBM, which validates both the discretization and the non-dimensionalization. Over the resolved range \(n\ge24\), the meshless scheme decays at approximately \(O(h^{2.7})\) and its error does not exceed that of the lattice baseline at equal resolution; at \(n=64\), for instance, the meshless error is \(1.49\times10^{-3}\), compared with \(3.32\times10^{-3}\) for the lattice scheme (at \(n=40\) the two are almost equal, \(4.45\times10^{-3}\) versus \(4.34\times10^{-3}\)). 

    In terms of physical recovery, the fitted viscosity ratio of the meshless method increases monotonically from \(\nu_{\mathrm{meas}}/\nu =0.9985\) at \(n=16\) to \(0.9992\) at \(n=64\), i.e.\ the measured viscosity exceeds the target by at most \(0.2\%\), indicating that the mimetic unitary streaming introduces a small interpolation-type numerical dissipation that vanishes with \(h\). The coarsest case \(n=8\) deviates from the asymptotic trend and is therefore excluded from the slope fit; its behavior is attributed to a pre-asymptotic, single-realization effect at low resolution. The lattice baseline reproduces the target viscosity to within \(0.04\%\) (ratios of \(0.9996\)-\(0.9997\)).The numerical results show that, the mass-weighted meshless full LBM reproduces the analytic TG decay with accuracy comparable to, and at finer resolutions better than, the regular-lattice LBM at equal physical parameters, while retaining viscosity recovery within \(0.1\%\).

\begin{figure}[htbp]
      \centering
      \includegraphics[width=1\linewidth]{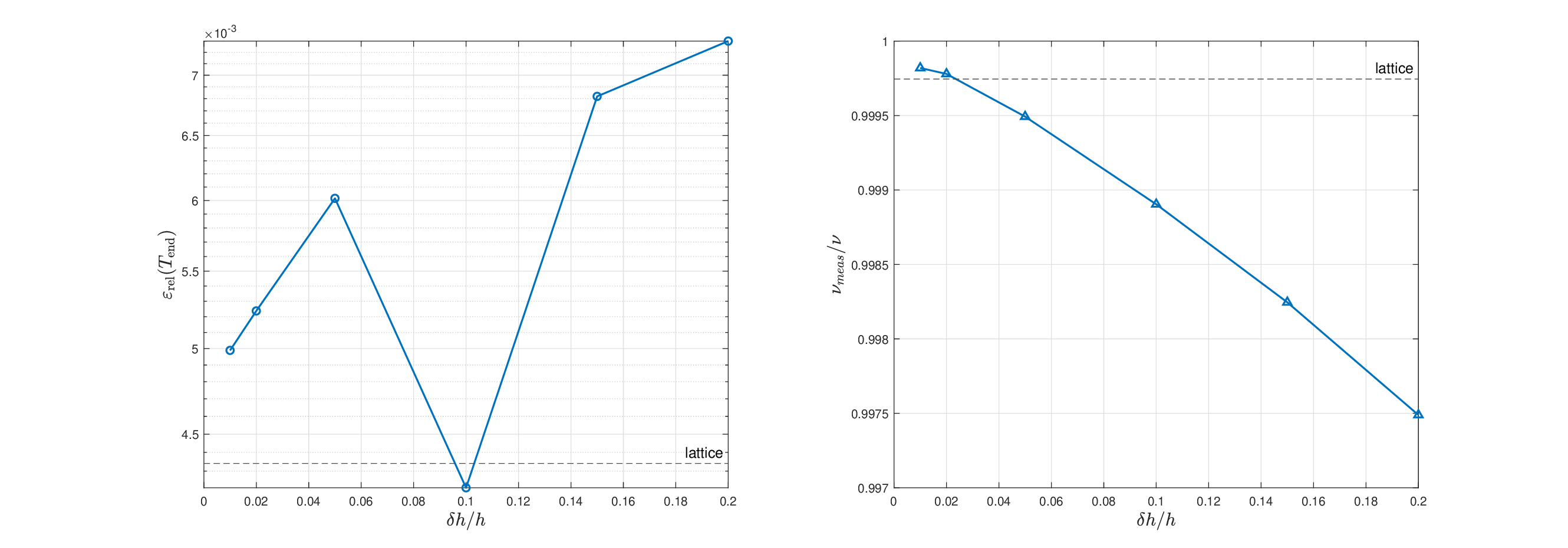}
      \caption{At a fixed resolution $n=40$, the accuracy of MW-MLBM after one advection time. Left: mass-weighted relative $L_2$ velocity error at the final time versus $\delta h$, with the black horizontal line denoting the corresponding error of the regular lattice LBM. Right: fitted viscosity ratio $\nu_{meas}/\nu$ versus $\delta h$, with the black horizontal line denoting the corresponding ratio of the regular lattice LBM.}
      \label{fig:MWMLBM-jit}
    \end{figure}

    We also investigated how the accuracy of the MW-MLBM method for simulating the Taylor–Green (TG) vortex flow varies with the degree of scatteredness of the point sets. The numerical results presented in Fig. \ref{fig:MWMLBM-jit} show that, the effect of point-cloud irregularity was examined at fixed resolution \(n=40\) with jitter amplitudes from \(0.01\) to \(0.20\), averaged over several realizations. The relative \(L_2\) error varies between \(4.2\times10^{-3}\) and \(7.3\times10^{-3}\) without a systematic trend, and is comparable to the lattice discretization error (\(\approx4.3\times10^{-3}\)) at this resolution, indicating that the overall accuracy is governed by the underlying LBM discretization rather than by the nodes irregularity. The viscosity ratio decreases slowly and monotonically from \(0.99981\) at jitter \(0.01\) to \(0.99748\) at jitter \(0.30\), i.e. the additional numerical dissipation grows from about \(0.019\%\) to \(0.25\%\) but remains below \(0.3\%\).

    Overall, the mass-weighted meshless LBM reproduces the Taylor–Green flow with accuracy comparable to, and at finer resolutions better than, the regular-lattice LBM, while recovering the target viscosity to within \(0.3\%\) over a wide range of points irregularity. These results establish that the full meshless scheme itself is feasible on scattered clouds and provide a reliable baseline for the Carleman-truncation study below.

  \subsection{Accuracy of the truncated Carleman linearized MW-MLBM}

    The Carleman linearization is an approximation of the discrete LBM evolution, not of the exact Navier-Stokes solution. We therefore assess the truncation error against the full mass-weighted meshless LBM computed with the same points set, the same mimetic unitary streaming operators, and the same collision model described in Eq.~\eqref{mass weighted collision}.

    The point cloud is a perturbed lattice on the periodic domain \([0,1)^2\) with jitter amplitude \(\delta h =0.1h\) and \(k=12\) nearest neighbors, so that each node carries its Voronoi control volume \(m_i\). The physical parameters are \(\mathrm{Re}=U_0L/\nu=50\) and \(U_0=0.05\) with \(L=1\), giving  \(\mathrm{Ma}=U_0/c_s\approx0.087\). For the resolution \(n=36\) (\(h=1/36\), \(N=1296\) nodes) the relaxation rate follows from $ \bar{\tau}=\frac12+ 3\nu/h$ and $\Omega=1/\bar{\tau}$. Each simulation advances one advection time, \(T_{\mathrm{end}}=L/U_0=20\), i.e.\ \(N_t=720\) time steps. 

\begin{figure}[htbp]
    \centering
      \includegraphics[width=1\linewidth]{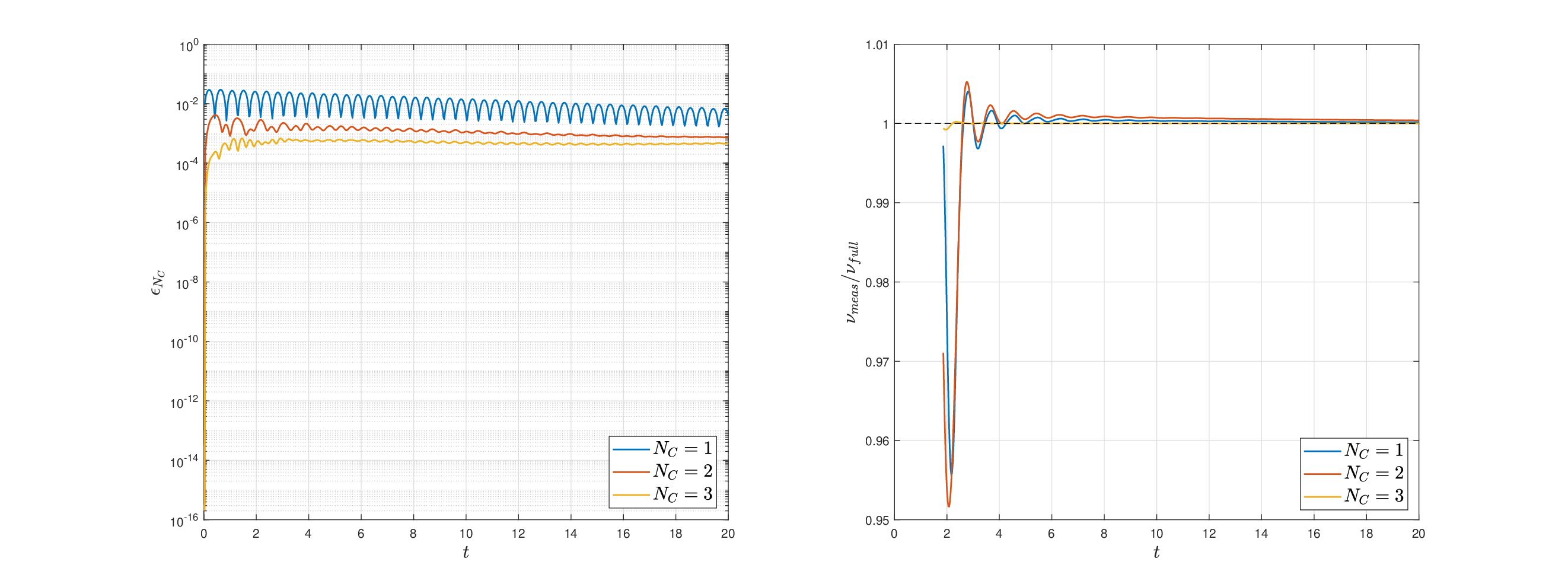}
      \caption{At a fixed resolution $n=36$, the accuracy of truncated Carleman systems with $N_C=1,2,3$ compared with full MW-MLBM. Left: mass-weighted relative $L_2$ velocity error versus time $t$. Right: fitted viscosity ratio $\nu_{\mathrm{meas}}^{N_C}(t)/\nu_{\mathrm{meas}}^{\mathrm{full}}(t)$ versus time $t$.}
      \label{fig:carleman-tgv}
    \end{figure}

    The Carleman systems are advanced with truncation orders \(N_C=1,2,3\) and compared with the full model through the mass-weighted relative \(L_2\) velocity error
   \begin{equation}
    \varepsilon_{N_C}(t) =\frac{\sqrt{\sum_i m_i\bigl|\boldsymbol u_i^{N_C}(t) -\boldsymbol u_i^{\mathrm{th}}(t)\bigr|^2}} {\sqrt{\sum_i m_i\bigl|\boldsymbol u_i^{\mathrm{th}}(t)
    \bigr|^2}},
    \end{equation}
    We also compare the measured viscosities of the truncated Carleman systems and full MW-MLBM. The effective viscosity is estimated cumulatively from the kinetic-energy decay: for every \(t\), \(\nu_{\mathrm{meas}}(t)\) is the slope of a linear fit of \(\ln E(t')\) over \(t'\in[t_0,t]\), with \(t_0=\sqrt3\) (approximately one acoustic period) 
    and the truncation is assessed through the ratio \(\nu_{\mathrm{meas}}^{N_C}(t)/\nu_{\mathrm{meas}}^{\mathrm{full}}(t)\). The numerical results are shown in Fig.~\ref{fig:carleman-tgv}.

    As shown in the left subplot, the truncation error decreases systematically with \(N_C\). The first-order truncation exhibits a mean deviation of \(1.03\times10^{-2}\) from the full model. Retaining the quadratic term (\(N_C=2\)) reduces the mean error to \(1.2 \times10^{-3}\), i.e.\ by a factor of \(8.4\). The cubic truncation (\(N_C=3\)) further reduces the mean error to \(4.8\times10^{-4}\), an additional factor of \(2.6\), so that the error is smaller than that of the first-order truncation by more than a factor of \(20\). In all cases the maximal error occurs during the initial transient, after which the error decays; the residual oscillatory component, associated with the acoustic modes excited by the truncated pressure-advection balance, is strongly suppressed at \(N_C=3\).

    The cumulative viscosity ratio is reported in the right subplot. At \(N_C=3\) the ratio agrees with the full model to within \(10^{-5}\) throughout the simulation, whereas the first- and second-order truncations differ by about \(10^{-3}\) on average. The initial sharp drop visible in the \(N_C=1,2\) curves (down to approximately \(0.95\)) is an acoustic transient: immediately after \(t_0\), the fitting interval \([t_0,t]\) is still short and the residual pressure imbalance of the truncated collision modulates the instantaneous decay rate. As the fitting interval grows, the cumulative estimate converges monotonically to unity.

    The non-monotonic ordering between \(N_C=1\) and \(N_C=2\) should not be interpreted as a loss of accuracy. The measured viscosity is a largely linear diagnostic: the leading dissipation is determined by the linear collision operator \(I+F_1\), which is common to all truncation orders, and \(\nu_{\mathrm{meas}}\) is therefore only weakly sensitive to \(N_C\). The quadratic term added at \(N_C=2\) redistributes energy among acoustic and higher-harmonic modes without providing the cubic feedback that first corrects the energy evolution; consequently it may shift the fitted decay rate slightly away from the full-model value, whereas the first-order truncation, being purely linear, remains accidentally close. At \(N_C=3\) the leading cubic correction is included and the agreement improves by two to three orders of magnitude. The velocity-field error, which decreases monotonically with \(N_C\), is therefore the appropriate metric for assessing the convergence of the Carleman truncation.

    Overall, for a fixed resolution and physical parameters, increasing the Carleman truncation order produces a systematic reduction of the velocity-field error, while the cumulative viscosity estimate converges to the full-model value, reaching agreement at the \(10^{-5}\) level at \(N_C=3\). The residual acoustic oscillations of the lower-order truncations are a consequence of the incomplete pressure-advection balance and diminish rapidly with \(N_C\), whereas their contribution to the cumulative viscosity estimate is averaged out by the running fit.

    \subsection{Normalization factor of the mass-weighting diagonal}

To quantify the subnormalization factor induced by the mass-weighting diagonal
\(M=\operatorname{diag}(m_i)\), we study the Voronoi cell areas of a jittered
periodic lattice. The effect of point-cloud irregularity was examined at fixed resolution \(n=40\) with jitter amplitudes from \(0\) to \(0.5\), averaged over 20 independent realizations.
For every realization we compute the toroidal Voronoi areas \(m_i\), normalize
\(\hat m_i=m_i/\bar m\), and record
\[
s_{\min}=\min_i\sqrt{\hat m_i},\qquad
\alpha_M=\parallel M^{-1/2}\parallel_{\infty}=\frac{1}{s_{\min}}.
\]

\begin{figure}[htbp]
  \centering
  \includegraphics[width=1\linewidth]{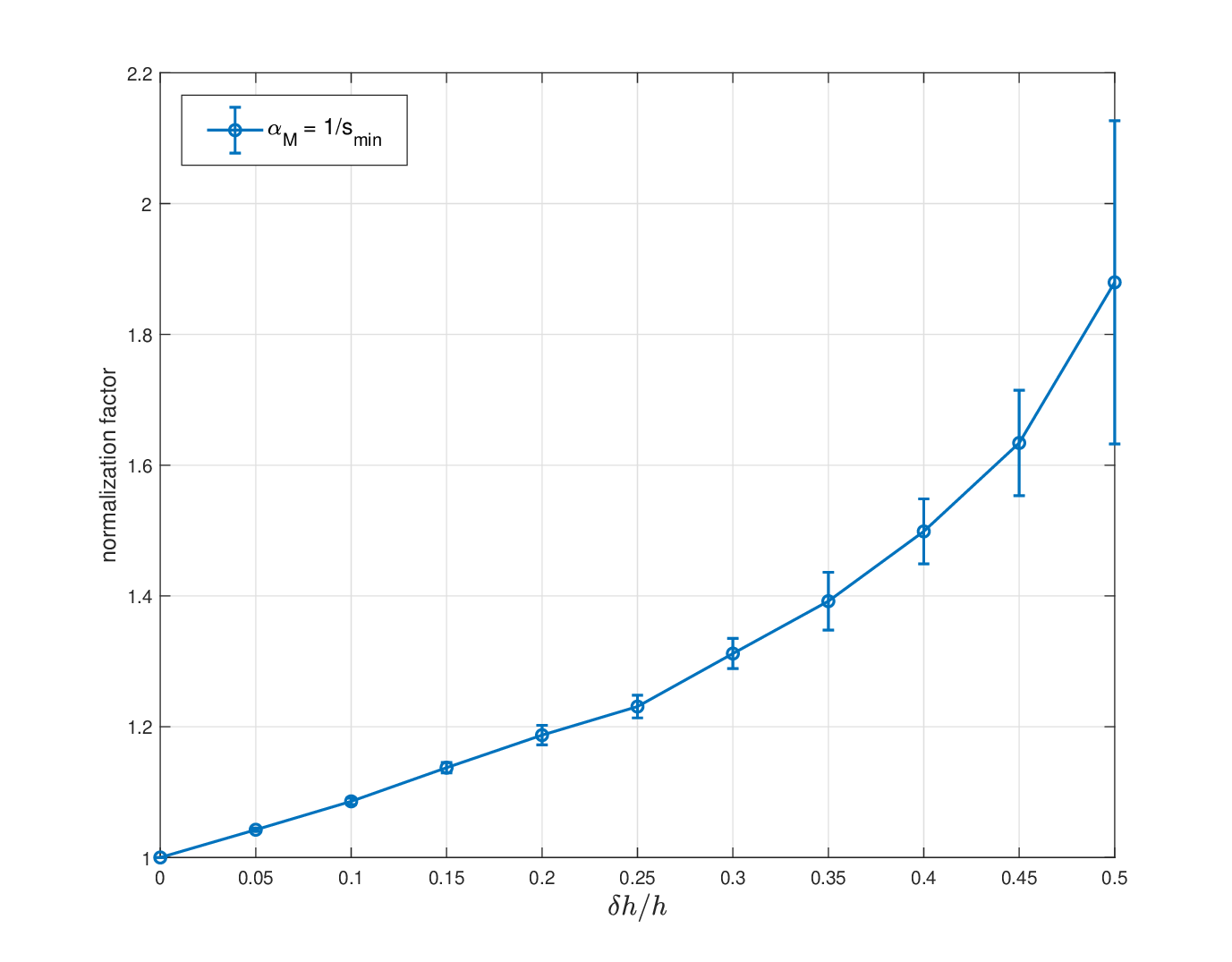}
  \caption{the normalization factor \(\alpha_M=1/s_{\min}\) as functions of the jitter amplitude, measured in units of the lattice spacing \(h\). Error bars denote one standard deviation over 20 realizations.}
  \label{fig:massweight-alpha}
\end{figure}

The data show a monotone decrease of \(s_{\min}\) from its optimal value
\(1\) on the uniform lattice to \(0.539\) at \(\delta h/h=0.5\), while
\(\alpha_M\) grows from \(1\) to \(1.88\). Throughout the practically relevant scatter range
\(\delta h/h\le 0.3\), \(\alpha_M\) remains below \(1.32\), confirming that
the mass-weighting normalization factor stays of order one. This behaviour is
consistent with the geometric lower bound
\(s_{\min}\ge\sqrt{\pi}\,(\tfrac12-\delta h/h)\): since the minimum pairwise
node distance is at least \((h-2\,\delta h)\), every Voronoi cell contains
a disk of radius \(\tfrac12(h-2\,\delta h)\), so the bound is independent
of \(N\). Only in the limit \(\delta h/h\to\tfrac12\), where neighboring
nodes may coalesce, does \(\alpha_M\) start to grow rapidly. Consequently the
block-encoding subnormalization of \(M^{-1/2}\), and with it the success probability
\(1/\alpha_M^{N_C}\) of the Carleman flow step, degrades only mildly for
quasi-uniform point sets.

\section{Conclusion}\label{section 7}

In this paper, we propose a quantum interpolation-supplemented meshless lattice Boltzmann method based on Carleman embedding. The probability density function was re-encoded through mass weighting, and a mimetic differential matrix was constructed as the generator of translation interpolation matrix, which is also the streaming operator. Its advantage is that the streaming step maintains the same unitary property as classical LBM. Meanwhile, we also demonstrate that the collision step changes caused by re-encoding have a minimal impact on the degradation of matrix block encoding complexity. Numerically, we validated the effectiveness of our proposed algorithm and its convergence with Carleman truncation order on Taylor Green problems with full periodic boundary conditions. Due to the requirement of unitary translation interpolation operator, our algorithm is suitable for full periodic boundary conditions, and further research on non-periodic boundary conditions problems will be conducted in the future.
\section*{Acknowledgements}\label{Acknowledgements}


\begin{thebibliography}{10}
\expandafter\ifx\csname url\endcsname\relax
  \def\url#1{\texttt{#1}}\fi
\expandafter\ifx\csname urlprefix\endcsname\relax\def\urlprefix{URL }\fi
\expandafter\ifx\csname href\endcsname\relax
  \def\href#1#2{#2} \def\path#1{#1}\fi

\bibitem{heluo1997}
X.~He, L.-S. Luo, A priori derivation of the lattice {B}oltzmann equation,
  Physical Review E 55~(6) (1997) R6333--R6336.
\newblock \href {https://doi.org/10.1103/PhysRevE.55.R6333}
  {\path{doi:10.1103/PhysRevE.55.R6333}}.

\bibitem{carleman1932}
T.~Carleman, Application de la th{\'e}orie des {\'e}quations int{\'e}grales
  lin{\'e}aires aux syst{\`e}mes d'{\'e}quations diff{\'e}rentielles non
  lin{\'e}aires, Acta Mathematica 59 (1932) 63--87.
\newblock \href {https://doi.org/10.1007/BF02546499}
  {\path{doi:10.1007/BF02546499}}.

\bibitem{kowalski1991}
K.~Kowalski, W.-H. Steeb, Nonlinear Dynamical Systems and Carleman
  Linearization, World Scientific, Singapore, 1991.

\bibitem{forets2017}
M.~Forets, A.~Pouly, Explicit error bounds for carleman linearization (2017).
\newblock \href {http://arxiv.org/abs/1711.02552} {\path{arXiv:1711.02552}}.

\bibitem{liu2021}
J.-P. Liu, H.~{\O}. Kolden, H.~K. Krovi, N.~F. Loureiro, K.~Trivisa, A.~M.
  Childs, Efficient quantum algorithm for dissipative nonlinear differential
  equations, Proceedings of the National Academy of Sciences 118~(35) (2021)
  e2026805118.
\newblock \href {https://doi.org/10.1073/pnas.2026805118}
  {\path{doi:10.1073/pnas.2026805118}}.

\bibitem{krovi2023}
H.~Krovi, Improved quantum algorithms for linear and nonlinear differential
  equations, Quantum 7 (2023) 913.
\newblock \href {https://doi.org/10.22331/q-2023-02-02-913}
  {\path{doi:10.22331/q-2023-02-02-913}}.

\bibitem{yepez2001}
J.~Yepez, Quantum lattice-gas model for computational fluid dynamics, Physical
  Review E 63~(4) (2001) 046702.
\newblock \href {https://doi.org/10.1103/PhysRevE.63.046702}
  {\path{doi:10.1103/PhysRevE.63.046702}}.

\bibitem{budinski2022}
L.~Budinski, Quantum algorithm for the {Navier--Stokes} equations by using the
  streamfunction--vorticity formulation and the lattice {B}oltzmann method,
  International Journal of Quantum Information 20~(2) (2022) 2150039.
\newblock \href {https://doi.org/10.1142/S0219749921500398}
  {\path{doi:10.1142/S0219749921500398}}.

\bibitem{sanavio2025}
C.~Sanavio, W.~A. Simon, A.~Ralli, P.~Love, S.~Succi,
  Carleman-lattice-{B}oltzmann quantum circuit with matrix access oracles,
  Physics of Fluids 37~(3) (2025) 037123.

\bibitem{jennings2025a}
D.~Jennings, K.~Korzekwa, M.~Lostaglio, R.~Ashworth, E.~Marsili, S.~Rolston, An
  end-to-end quantum algorithm for nonlinear fluid dynamics with bounded
  quantum advantage (2025).
\newblock \href {http://arxiv.org/abs/2512.03758} {\path{arXiv:2512.03758}}.

\bibitem{jennings2025b}
D.~Jennings, K.~Korzekwa, M.~Lostaglio, P.~Mannix, R.~Ashworth, E.~Marsili,
  S.~Rolston, Simulating non-trivial incompressible flows with a quantum
  lattice {B}oltzmann algorithm (2025).
\newblock \href {http://arxiv.org/abs/2512.05781} {\path{arXiv:2512.05781}}.

\bibitem{heluodembo1996}
X.~He, L.-S. Luo, M.~Dembo, Some progress in lattice {B}oltzmann method. {Part
  I.} nonuniform mesh grids, Journal of Computational Physics 129~(2) (1996)
  357--363.
\newblock \href {https://doi.org/10.1006/jcph.1996.0255}
  {\path{doi:10.1006/jcph.1996.0255}}.

\bibitem{shu2002}
C.~Shu, X.~D. Niu, Y.~T. Chew, Taylor-series expansion and least-squares-based
  lattice {B}oltzmann method: two-dimensional formulation and its applications,
  Physical Review E 65~(3) (2002) 036708.
\newblock \href {https://doi.org/10.1103/PhysRevE.65.036708}
  {\path{doi:10.1103/PhysRevE.65.036708}}.

\bibitem{strzelczyk2023}
D.~Strzelczyk, M.~Matyka, Study of the convergence of the meshless lattice
  {B}oltzmann method in {Taylor--Green} and annular channel flows (2023).
\newblock \href {http://arxiv.org/abs/2306.01525} {\path{arXiv:2306.01525}}.

\bibitem{fornberg2015}
B.~Fornberg, N.~Flyer, A Primer on Radial Basis Functions with Applications to
  the Geosciences, SIAM, Philadelphia, 2015.
\newblock \href {https://doi.org/10.1137/1.9781611974041}
  {\path{doi:10.1137/1.9781611974041}}.

\bibitem{lowchuang2019}
G.~H. Low, I.~L. Chuang, Hamiltonian simulation by qubitization, Quantum 3
  (2019) 163.
\newblock \href {https://doi.org/10.22331/q-2019-07-12-163}
  {\path{doi:10.22331/q-2019-07-12-163}}.

\bibitem{gilyen2019}
A.~Gily{\'e}n, Y.~Su, G.~H. Low, N.~Wiebe, Quantum singular value
  transformation and beyond: exponential improvements for quantum matrix
  arithmetics, in: Proceedings of the 51st Annual ACM SIGACT Symposium on
  Theory of Computing (STOC), 2019, pp. 193--204.
\newblock \href {https://doi.org/10.1145/3313276.3316366}
  {\path{doi:10.1145/3313276.3316366}}.

\bibitem{lloyd1996}
S.~Lloyd, Universal quantum simulators, Science 273~(5278) (1996) 1073--1078.
\newblock \href {https://doi.org/10.1126/science.273.5278.1073}
  {\path{doi:10.1126/science.273.5278.1073}}.

\bibitem{shashkov1996}
M.~Shashkov, Conservative Finite-Difference Methods on General Grids, CRC
  Press, Boca Raton, 1996.

\bibitem{lipnikov2014}
K.~Lipnikov, G.~Manzini, M.~Shashkov, Mimetic finite difference method, Journal
  of Computational Physics 257 (2014) 1163--1227.
\newblock \href {https://doi.org/10.1016/j.jcp.2013.07.031}
  {\path{doi:10.1016/j.jcp.2013.07.031}}.

\bibitem{desbrun2005}
M.~Desbrun, A.~N. Hirani, M.~Leok, J.~E. Marsden, Discrete exterior calculus
  (2005).
\newblock \href {http://arxiv.org/abs/math/0508341}
  {\path{arXiv:math/0508341}}.

\bibitem{xiao2026}
Y.~Xiao, L.~Yang, H.~Dong, Y.~Du, Interpolation-supplemented quantum lattice
  {B}oltzmann method for incompressible flows with nonuniform mesh and curved
  boundary, Physical Review E 113 (2026) 045306.
\newblock \href {https://doi.org/10.1103/PhysRevE.113.045306}
  {\path{doi:10.1103/PhysRevE.113.045306}}.

\bibitem{yang2026}
L.~Yang, Y.~Xiao, C.~Shu, Y.~Du, Quantum {T}aylor series expansion and least
  squares-based fractional-step {LBM} for incompressible flows on non-uniform
  meshes, Chinese Journal of Aeronautics 39~(6) (2026) 104176.
\newblock \href {https://doi.org/10.1016/j.cja.2026.104176}
  {\path{doi:10.1016/j.cja.2026.104176}}.

\bibitem{Xiang2022}
H.~Xiang, Quantum Numerical Linear Algebra, Tsinghua University Press, Beijing,
  2022, in Chinese.

\bibitem{Dalzell2024}
A.~Dalzell, A shortcut to an optimal quantum linear system solver (06 2024).
\newblock \href {https://doi.org/10.48550/arXiv.2406.12086}
  {\path{doi:10.48550/arXiv.2406.12086}}.

\bibitem{suzuki1990}
M.~Suzuki, Fractal decomposition of exponential operators with applications to
  many-body theories and monte carlo simulations, Physics Letters A 146~(6)
  (1990) 319--323.
\newblock \href {https://doi.org/10.1016/0375-9601(90)90962-N}
  {\path{doi:10.1016/0375-9601(90)90962-N}}.

\end{thebibliography}
\end{document}